\documentclass[aps,prb,twocolumn,superscriptaddress]{revtex4-1}

\usepackage{graphicx}
\usepackage{amsmath}
\usepackage{bm}
\usepackage{xcolor}
\usepackage{multirow}
\begin{document}


\title{Order-disorder transition in the prototypical antiferroelectric PbZrO$_3$}


\author{Bin Xu}
\email[Email address: ]{xubin.physics@gmail.com}
\affiliation{School of Physical Science and Technology, Soochow University, Suzhou 215006, China}
\affiliation{Physics Department and Institute for Nanoscience and Engineering, University of Arkansas, Fayetteville, Arkansas 72701, USA}

\author{Olle Hellman}
\affiliation{Department of Physics, Chemistry and Biology (IFM), Link\"oping University, Link\"oping SE-581 83, Sweden}

\author{L. Bellaiche}
\email[Email address: ]{laurent@uark.edu}
\affiliation{Physics Department and Institute for Nanoscience and Engineering, University of Arkansas, Fayetteville, Arkansas 72701, USA}


\date{\today}

\begin{abstract}
The prototypical antiferroelectric PbZrO$_3$ has several unsettled questions, such as the nature of the antiferroelectric transition, possible intermediate phase and the microscopic origin of the \textit{Pbam} ground state. Using first principles, we show that no phonon becomes truly soft at the cubic-to-\textit{Pbam} transition temperature, and the order-disorder character of this transition is clearly demonstrated based on molecular dynamics simulations and potential energy surfaces. The out-of-phase octahedral tilting is an important degree of freedom, which can collaborate with other phonon distortions and form a complex energy landscape with multiple minima. Candidates of the possible intermediate phase are suggested based on the calculated kinetic barriers between energy minima, and the development of a  first-principles-based effective Hamiltonian. The use of this latter scheme further reveals that specific {\it bi-linear} interactions between local dipoles and octahedral tiltings  play a major role in the formation of the \textit{Pbam} ground state, which contrasts with most of the previous explanations.
\end{abstract}


\maketitle


Antiferroelectrics (AFEs) form a special class of materials that possess anti-polar displacements, which was first conceptualized by Kittel \cite{kittel1951}. AFEs have received much attention in recent years because they hold great potential to reach high energy density for high-power energy storage \cite{hao2013,rabe2013,xu2017}. Lead zirconate PbZrO$_3$ (PZO) is the first discovered AFE material, and crystallizes in a \textit{Pbam} ground state below $\sim$505 K \cite{sawaguchi1951,shirane1951,shirane1952}. Besides the anti-polar distortions, this particular structure also exhibits strong long-ranged order  tiltings of oxygen octahedra, which is often termed as antiferrodistortive (AFD) distortions.

Despite the efforts devoted to understanding the origin of the AFE ground state in PZO \cite{tagantsev2013,jorge2014,hlinka2014}, how distortions condense from the high-symmetry cubic structure remains an open issue. Unlike \textit{proper} ferroelectrics where the ferroelectric (FE) distortion results from soft polar mode in the cubic phase, it is presently unclear if PZO is a \textit{proper} AFE.  As a matter of fact, on one hand, completely soft AFE and AFD modes  have been predicted at the AFE transition temperature ($T_A$) by Fthenakis \textit{et al.} based on an effective Hamiltonian method  \cite{fthenakis2017}. On the other hand, no truly soft AFE or AFD modes was observed in Ref. \cite{tagantsev2013}. In this regards, Tagantsev \textit{et al}. proposed that the AFE transition in PZO is driven by a non-critical softening of the transverse acoustic mode via flexoelectric coupling \cite{tagantsev2013}; however, this mechanism is rather exotic and has recently been ruled out \cite{vales2018}. Moreover, from an energetic point of view, \'I\~niguez \textit{et al.}. believe that the AFD mode plays a crucial role and PZO may be an \textit{improper} AFE \cite{jorge2014}.

In fact, the above considerations in Refs.~\cite{fthenakis2017,jorge2014} rely on the existence of a displacive transition, for which the distortions in the antiferroelecrtic state are derived from sudden deviations from the ideal cubic structure at $T_A$. However, such assumption may be invalid for PZO, as the AFE transition have been advocated to possess a strong order-disorder character, as evidenced by (i) the existence of central mode \cite{ostapchuk2001,ko2013} and (ii) the fact that disordered Pb displacement and oxygen tiltings were found to occur even in the cubic phase \cite{pasciak2015,zhang2015}. Interestingly, this picture of dynamic disorder in the cubic phase is similar to the recent studies of the AFD transitions in CaMnO$_3$ and inorganic halide perovskites \cite{klarbring2018,klarbring2019}, where low-energy paths between local minima are found to be critical for understanding the finite-temperature transitions.

Furthermore, there exists a FE \textit{R3c} phase that is energetically very close to the AFE state, merely a few meV/f.u. higher than \textit{Pbam} at 0 K, according to \textit{ab initio} calculations \cite{sebastian2013,jorge2014}. Hence, it is interesting to understand what microscopic effect is mostly responsible for the stabilization of \textit{Pbam} rather than \textit{R3c} as the ground state. In particular, is it the commonly believed trilinear energy coupling \cite{hlinka2014,jorge2014}, or rather something else \cite{patel2016}?  Another unsettled issue in PZO concerns the possible existence of an intermediate phase in the vicinity of $T_A$ \cite{sawaguchi1951,tennery1965,tennery1966,goulpeau1967,fujishita1992,liu2018}: such intermediate state can only occur over a very narrow temperature range of 3--5 K between the AFE and the cubic phase, and its structure remains a mystery.

In this article, we use first-principles methods to demonstrate that the order-disorder picture may hold the key to understand the AFE transition and the debated intermediate phase \cite{sawaguchi1951,tennery1965,tennery1966,goulpeau1967,fujishita1992}  in PZO. We also reveal that a very specific bi-linear (and therefore not trilinear) energetic coupling plays  a crucial role in the stabilization of \textit{Pbam} and can also yield other more complex structures, such as with long-periods and incommensurable ones.



Phonons in the cubic phase at high temperatures are investigated by combining \textit{ab initio} molecular dynamics (AIMD) simulations implemented in \textsc{vasp} \cite{kresse1999} with the temperature dependent effective potential (TDEP) method \cite{hellman2011,hellman2013,hellman2013_2}. The potential energy surface (PES) is calculated by density functional theory (DFT) \cite{kresse1999}, and the kinetic barriers are computed with the climbing image nudged elastic band (CI-NEB) method \cite{henkelman2000}. We also develop an effective Hamiltonian for PZO to explain the microscopic origin of its complex ground state and to suggest other more complex candidates for the intermediate state. Further details of the methods are given in the Supplemental Material (SM) \cite{sm}.

\begin{figure}[thb]
	\includegraphics[totalheight=0.17\textheight]{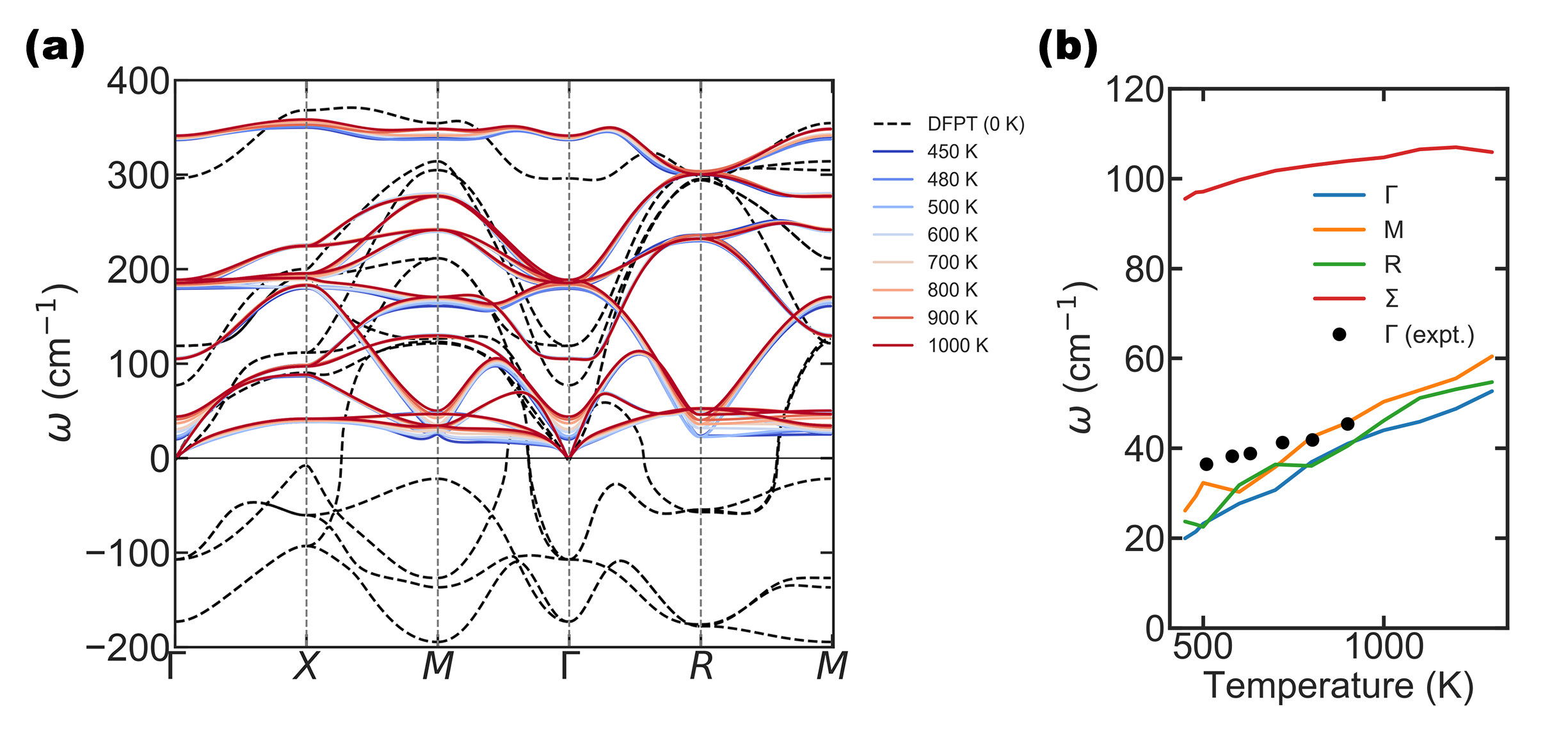}
	\caption{Calculated temperature dependence of phonon frequencies in the cubic phase. (a) Phonon dispersion spectra at various temperatures. For clarity, only branches with $\omega<$ 400 cm$^{-1}$ are shown. For comparison, DFPT phonon is also displayed (dashed lines). (b) Temperature dependence of the frequency of the ferroelectric mode ($\Gamma$), in-phase tilting mode ($M$), out-of-phase tilting mode ($R$), and anti-polar mode ($\Sigma$), in comparison with available experimental data \cite{ostapchuk2001}.} \label{fig1}
\end{figure}

A purely displacive picture requires truly soft phonon mode. Let us first check how phonon softens in the cubic phase as temperature decreases. The dispersion for different temperatures above 450 K are shown in Fig. \ref{fig1}a, as calculated by the TDEP method based on AIMD simulations with a 4$\times$4$\times$4 supercell (320 atoms). In contrast to the 0K phonon computed by density functional perturbation theory (DFPT) that exhibits soft branches \cite{ghosez1999,cockayne2000,leung2002,leung2003} (also shown in Fig. \ref{fig1}a via dashed lines), all the phonon frequencies are positive for $T>$ 450 K, implying that the cubic structure is dynamically stable at high temperature. We also numerically find via AIMD simulations  that, below 450 K, the structure is not cubic \cite{note1}. Interestingly, this latter temperature is rather close to the experimental AFE transition temperature $T_A$ of $\sim$505 K. 

Moreover, the 0K phonon exhibits the largest instabilities at the $\Gamma$, $R$, and $M$ points of the Brillouin zone, that correspond to $\bm{q}_{\Gamma}=(2\pi/a)(0,0,0)$, $\bm{q}_{R}=(2\pi/a)(1/2,1/2,1/2)$, and $\bm{q}_{M}=(2\pi/a)(1/2,1/2,0)$, respectively (with $a$ being the 5-atom lattice constant) and that are associated with  the FE distortion, out-of-phase (\textit{oop}) tilting, and in-phase (\textit{ip}) tilting, respectively. Furthermore, according to the calculated dispersion curves, when cooling from high temperature, these phonon modes show noticeable softening, as depicted in Fig. \ref{fig1}b. Note that the calculated zone-center softening agrees reasonably well with the measured data for the $\Gamma$-point mode \cite{ostapchuk2001}.

Interestingly, no phonon becomes truly soft (that is none of them drops to zero frequency) when the temperature approaches $\simeq$ 450K from above, as consistent with experimental data in the cubic phase \cite{tagantsev2013,ostapchuk2001}. In particular, Ref. \cite{tagantsev2013} pointed out that the $\Sigma$- and $R$-point modes are not soft at $T_A$, where $\bm{q}_{\Sigma}=(2\pi/a)(1/4,1/4,0)$, which is consistent with our prediction in Fig. \ref{fig1}b, but contrasts with the theoretical work of Ref. \cite{fthenakis2017} that predicted completely softened $\Sigma$ and $R$ modes.

\begin{figure}[thb]
	\includegraphics[totalheight=0.3\textheight]{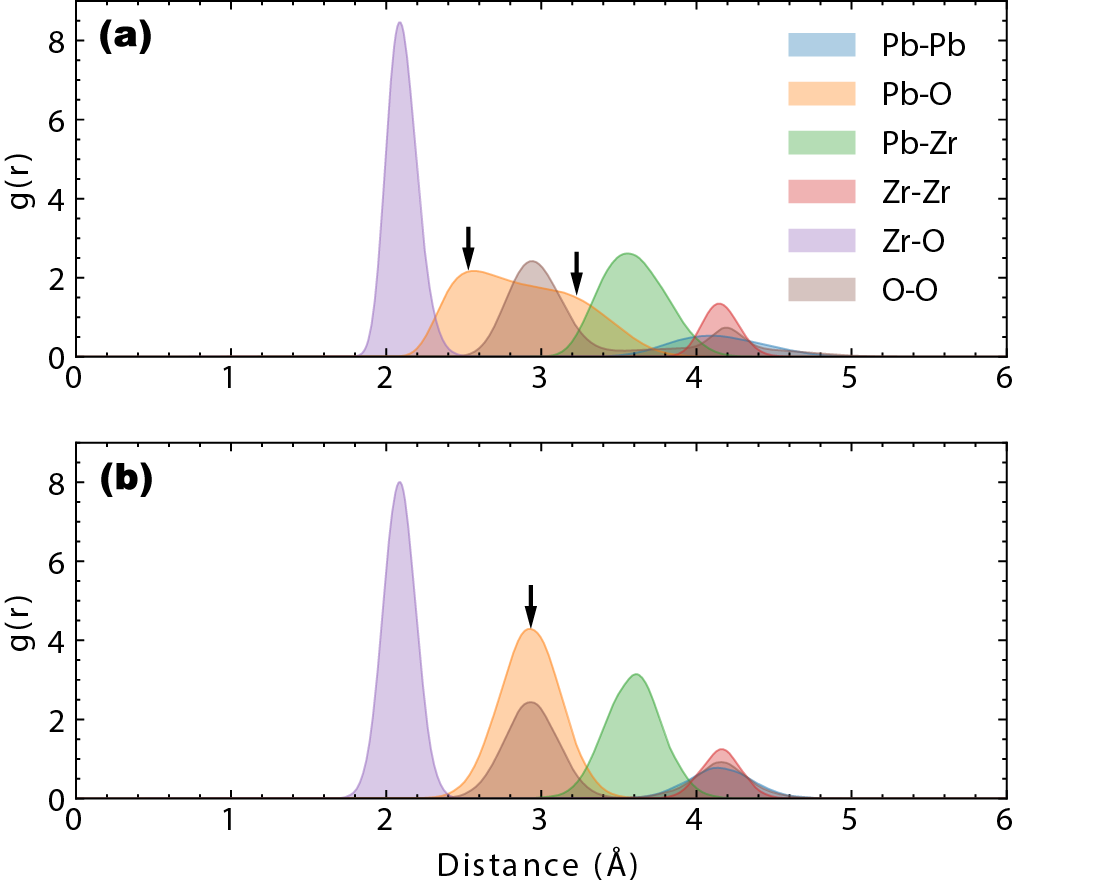}
	\caption{Pair correlation function as an indication of structural distortions at 600 K. (a) \textit{Ab initio} MD sampling. (b) Stochastic sampling. Black arrows point to the differences.} \label{fig2}
\end{figure}

The lack of soft phonon suggests that the paraelectric-to-AFE transition (without counting the intermediate phase) is of first order,  as also consistent with experiments \cite{samara1970,whatmore1979,fujishita2003}. Moreover, in order to check if  this transition has characters of an order-disorder type, as suggested by Ref. \cite{ostapchuk2001} based on the measured central mode, we analyze the MD simulation at 600 K, for which the cubic phase is stable. Figure \ref{fig2}a reports the pair correlation function $g(r)$, which is related to the probability of finding an ion at a given distance from another ion. For comparison, we also plot $g(r)$ from cubic stochastic sampling in Fig. \ref{fig2}b, in which the atomic displacements are centered harmonically around the cubic positions, as indicated by the Gaussian distribution of each pair. The only obvious difference between the two sampling schemes lies in the Pb-O pair, which has a single peak with cubic stochastic sampling but splits into two distances with MD sampling. This splitting is due to the finite tilting angle as long-life-time positions in the cubic phase, as evidenced from the MD simulation (see Figs. S2 and S3 in the SM \cite{sm} for more information). As the AIMD simulation is considered to be much more realistic than cubic stochastic sampling  \cite{note3}, it indicates that the cubic phase should be regarded as a {\it dynamical average structure}, in which thermal excitations cause the system to hop among local minima with finite tilting angles, with low-energy paths that bypass the ideal cubic positions (see Fig. S4 in the SM \cite{sm}). Such splitting also exists at even higher temperature, e.g., up to 1000 K, except that $g(r)$ gradually becomes closer to the cubic stochastic case when increasing $T$.

\begin{figure}[thb]
	\includegraphics[totalheight=0.23\textheight]{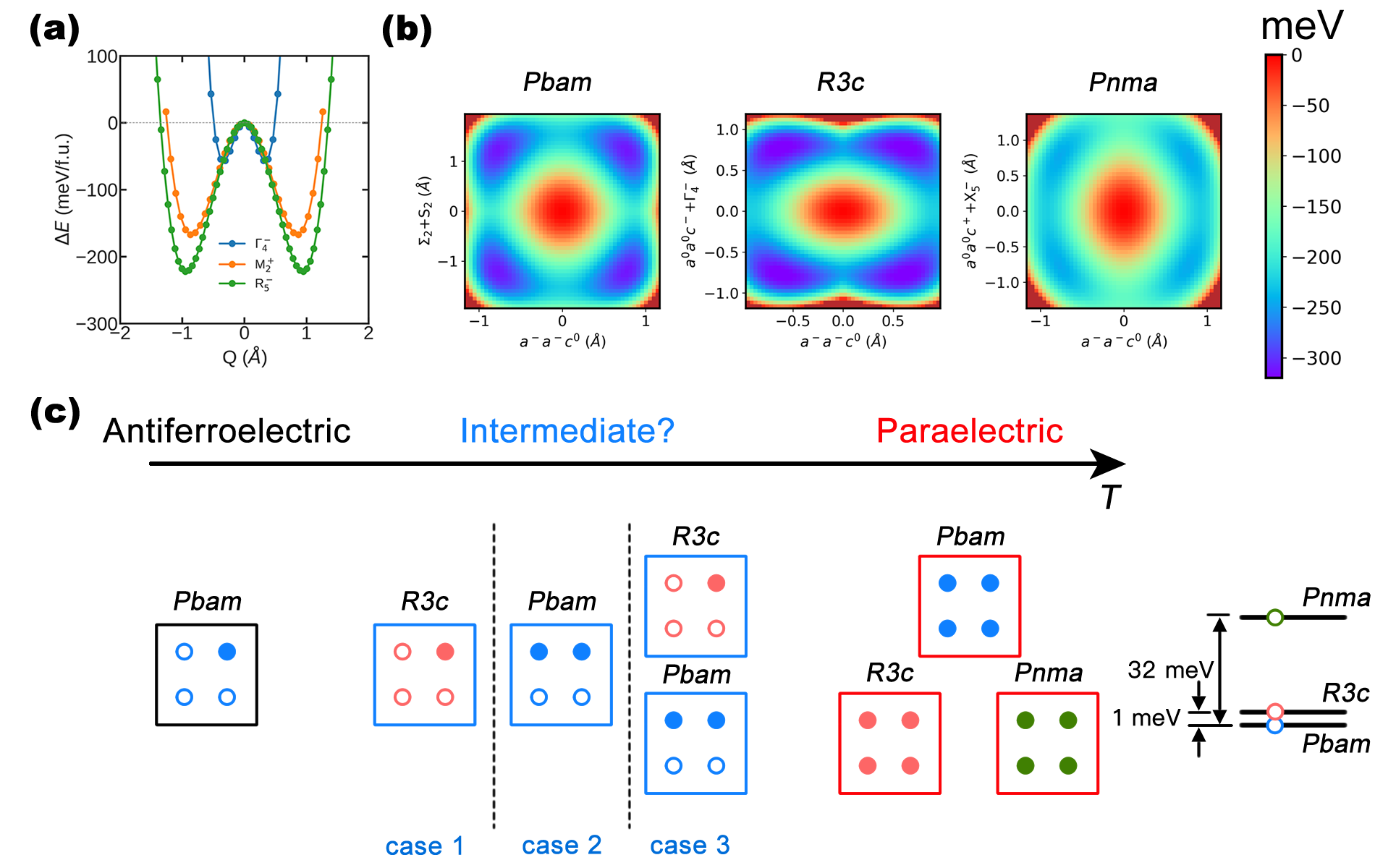}
	\caption{Potential energy landscapes in PZO. (a) 0K potential energy as a function of the amplitude of unstable high-symmetry phonon modes, i.e., $\Gamma$-point ferroelectric distortion, $M$-point in-phase tilting, and $R$-point out-of-phase tilting. (b) 0K PES as functions of $a^-a^-c^0$ and $\Sigma_2$+S$_2$ (left panel, \textit{Pbam} as minima), $a^-a^-c^0$ and $a^0a^0c^-$+$\Gamma^-_4$ (middle panel, \textit{R3c} as minima), and $a^-a^-c^0$ and $a^0a^0c^+$+X$^-_5$ (right panel, \textit{Pnma} as minima). (c) Simplified picture of the dynamic average structures. The circles in each square (\textit{Pbam}, \textit{R3c}, or \textit{Pnma} phase) denote the equivalent energy minima. The filled circles represent the minima being dynamically sampled at finite temperature. Note that four degeneracies chosen in this schematization is for drawing convenience and does not correspond to the real degeneracy of each phase.} \label{fig3}
\end{figure}

Further understanding of the structures and nature of the transition can be obtained from the PES. With the cubic structure as reference, we plot the double well potential with respect to a single mode distortion (of normalized magnitude denoted as $Q$), as shown in Fig. \ref{fig3}a.  The energy can be lowered with finite FE distortion ($\Gamma^-_4$), \textit{ip} tilting ($a^0a^0c^+$ or $M^+_2$), or \textit{oop} tilting ($a^-a^-c^0$ or $R^-_5$), which are the most significant instabilities in cubic PZO (see DFPT phonon in Fig. \ref{fig1}a). The largest energy gain of 222 meV/f.u. is provided by the $R^-_5$ mode, whereas the $M^+_2$ and $\Gamma^-_4$ modes cause less energy change, i.e., 167 and 57 meV/f.u., respectively. Such result implies that the \textit{oop} tilting is likely to be the most important degree of freedom in PZO. This is in line with the work of \'I\~niguez \textit{et al.}, which suggests that this mode plays a crucial role in stabilizing the AFE ground state \cite{jorge2014}. It is also worth to mention that all the low-energy structures reported in Ref. \cite{jorge2014} (e.g., \textit{Pbam}, \textit{R3c}, \textit{Pnma}, etc.) have the $a^-a^-c^0$ tilting as a component, further reflecting its role leading to low-energy states. Nevertheless, the large barrier heights of reversing $R^-_5$ and $M^+_2$ imply that the transition is not likely to be displacive, as thermal energy at $T_A$ is not sufficient to bypass the ideal cubic positions, even for condensing a single tilting mode. Such fact is related to the finite tilting angles in the cubic phase from MD simulations.

Next, we combine $a^-a^-c^0$ tilting with other distortions to yield three competing polymorphs having low energies, viz., the \textit{Pbam} (ground state), \textit{R3c}, and \textit{Pnma} phases. Using the crystallographic tool \textsc{amplimode} \cite{orobengoa2009,perez2010}, the \textit{Pbam} structure is mainly composed of three distortions: $R^-_5$, $\Sigma_2$ and S$_2$. The $R^-_5$ component captures 58.8\% of the total distortion, which is the largest, while it is 37.1\% and only 3.8\% for $\Sigma_2$ and S$_2$, respectively. To simplify the picture, we combine the $\Sigma_2$ and S$_2$ modes and compute the 2-dimensional PES by changing $R^-_5$ and $\Sigma_2$+S$_2$. The resulting contour plot is shown in the left panel of Fig. \ref{fig3}b, and exhibits four minima that correspond to the \textit{Pbam} structure \cite{note2} -- which is a manifestation of collaborative couplings between the $R^-_5$, $\Sigma_2$ and S$_2$ modes. The energy of these minima relative to the cubic structure (center) is 311 meV/f.u., being significantly larger than the barrier heights between these minima, which are found to be 89 or 113 meV/f.u. by reversing $R^-_5$ or $\Sigma_2$+S$_2$, respectively. Improved estimation of the barrier heights are obtained by the CI-NEB method, yielding 51 or 58 meV/f.u., respectively. Considering the energy scale of thermal energy (e.g., $k_BT=$ 86 meV at 1000 K), it appears unlikely that the ideal cubic structure is visited even at very high temperature, therefore further implying an order-disorder nature of the paraelectric-to-AFE transition (intermediate phase not counted).

The middle (respectively, right) panel of Fig. \ref{fig3}b depicts the PES associated with the changes of $R^-_5$ and  $a^0a^0c^-$+$\Gamma^-_4$ (respectively,  of $R^-_5$ and $a^0a^0c^+$+X$^-_5$), and therefore involves the \textit{R3c} (respectively,  \textit{Pnma}) phase. The coupling between $R^-_5$　and $a^0a^0c^-$+$\Gamma^-_4$ (or $a^0a^0c^+$+X$^-_5$) is also collaborative, and there exist low-barrier energetic paths without passing through the cubic structure too, which confirms that the high-temperature cubic phase should be taken as a {\it dynamical} average structure, with the system hopping between different states with finite AFD tiltings. This finding is in full accordance with the pair correlation function obtained from MD simulations (Fig. \ref{fig2}a).

Combining $a^-a^-c^0$ tilting with other phonon modes can therefore lead to these three low-energy structures, the ground state being \textit{Pbam} (set as $\Delta E=$ 0), the \textit{R3c} phase with an energy that is only 1 meV/f.u. higher, and $\Delta E=$ 32 meV/f.u. for the \textit{Pnma} phase. We will discuss the couplings that gives rise to the AFE ground state later in the text, but here one should note that the lower energy of \textit{Pbam} than \textit{R3c} depends on the precise structure and chemistry. Slight change of structure or chemical composition (e.g., Ti doping) can affect the energy hierarchy, giving rise to a FE \textit{R3c} ground state \cite{ayyub1998,boldyreva2007,mani2015,woodward2005}.

Although the \textit{R3c} and \textit{Pnma} phases do not occur in bulk PZO at any temperature, their minima are likely to be visited near $T_A$ or above. To see this, we calculate the 0K barrier from \textit{Pbam} to \textit{R3c} by the CI-NEB method, which is 57 meV/f.u., and the barrier is 53 mev/f.u. from \textit{Pbam} to \textit{Pnma}. Interestingly, the barriers between equivalent \textit{Pbam} structures is of similar magnitudes, i.e., 51 (or 58) meV/f.u. for reversing $R^-_5$ (or $\Sigma_2$+S$_2$).

This explains the occurrence of the intermediate phase in a very narrow temperature range, and the sensitive dependence on impurities, domain orientation, etc. \cite{tennery1965,tennery1966,goulpeau1967,liu2018}, as it only occurs when the system samples a fraction of the energy minima. It also brings to different possible scenario regarding the possible intermediate structure near the AFE transition temperature $T_A$ (Fig. \ref{fig3}c). For instance, the \textit{R3c} state is energetically very close to \textit{Pbam} at 0K,  and may even have a free energy near $T_A$ that is slightly lower than \textit{Pbam}, which would suggest that the  \textit{R3c} state can be the intermediate structure -- which is in line with the ferroelectric intermediate phase with rhombohedral symmetry that was observed in some experimental studies \cite{tennery1965,tennery1966,goulpeau1967,scott1972,whatmore1979}. Moreover, the  hopping between the minima of \textit{Pbam} of opposite \textit{oop} tiltings has rather lower barrier (51 meV/f.u., see left panel of Fig. \ref{fig3}b), which can imply that the intermediate state is an average state of \textit{Pbam} symmetry too but that does not possess any time-averaged finite $a^-a^-c^0$ tilting. However, as no orthorhombic intermediate phase has been reported in experiment, this latter \textit{Pbam} phase probably does not exist. Another possibility for the intermediate state is a more complex case  consisting of an average between  \textit{R3c} and some (or all)  \textit{Pbam} equivalent minima, reflecting the possible hoppings between all these states. Note that the difficulty of determining the symmetry of the intermediate phase in experiments may reflect such picture, i.e., PZO dynamically samples multiple local minima near $T_A$. On the other hand, we do not expect the  \textit{Pnma} state to be (significantely) involved in the intermediate state because its 0K energy is 32 meV/f.u. higher than \textit{Pbam}.

\begin{figure}[t]
	\includegraphics[totalheight=0.15\textheight]{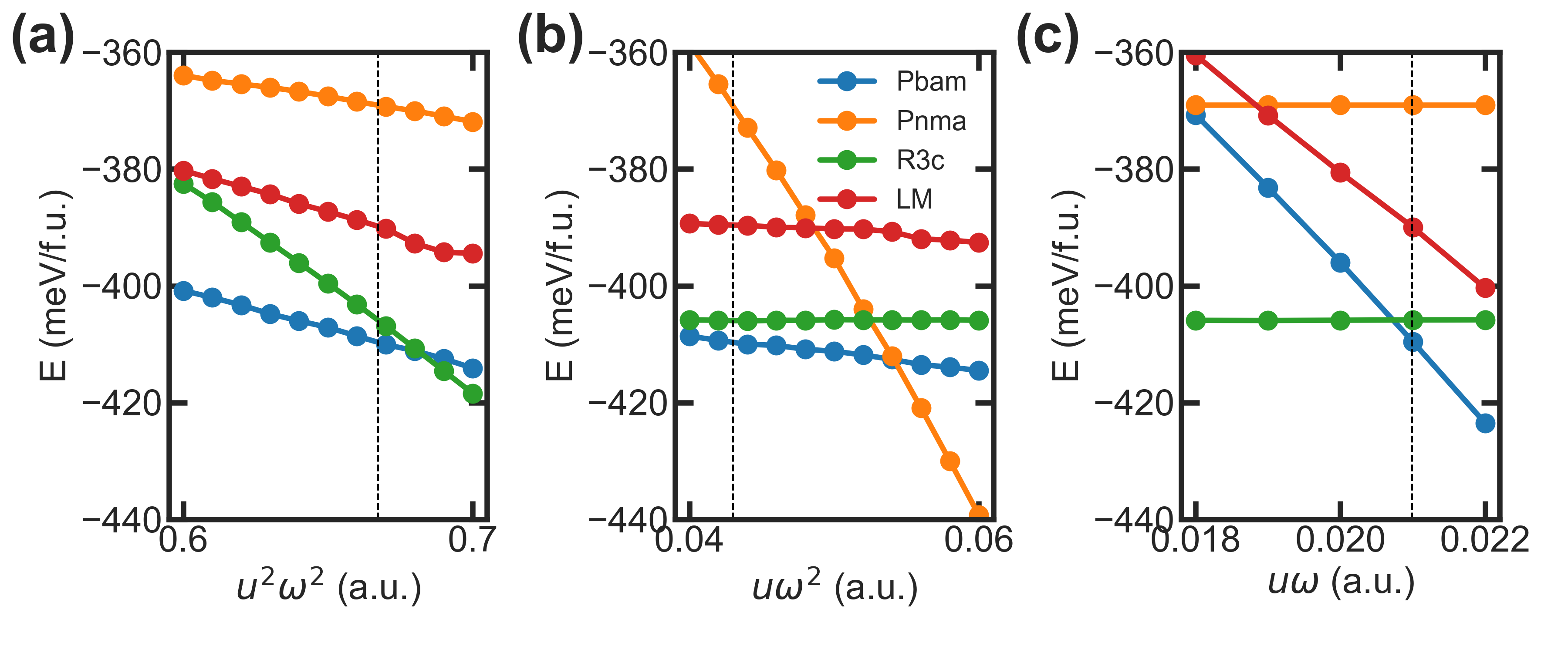}
	\caption{The effect of different types of dipole-tilting couplings on the total energies of the \textit{Pbam}, \textit{Pnma}, \textit{R3c}, and long-modulated (LM) phases of PbZrO$_3$ based on Monte Carlo simulations at 1 K. The simulated LM  phase corresponds to $\bm{q}_{\text{LM}}=(2\pi/a)(3/16,3/16,0)$ (a) Bi-quadratic coupling. (b) Tri-linear coupling. (c) Bi-linear coupling. The vertical dashed lines denote the coupling constants obtained by DFT calculations.} \label{fig4}
\end{figure}

So far we have demonstrated that deep local minima and the existence of low-barrier energetic paths bypassing the cubic positions are crucial to understand the finite-temperature structures in PZO. Now, we want to add some insight about why the \textit{Pbam} phase is the ground state of PZO, instead of \textit{R3c} or \textit{Pnma}, and suggest other possible intermediate states that are too complex to be treated by direct first-principles techniques. To this end, we developed an effective Hamiltonian for PZO, which includes energy terms involving local modes (proportional to the local electric dipole), local oxygen octahedral tiltings, strain, and the coupling among them (see SM for more details \cite{sm}); In particular, the three energy terms related to the couplings between the local dipoles and AFD distortions are given by:
\begin{align}\label{eq:afd}
\Delta E^{\text{AFD}}=&\sum_{ij}\sum_{\alpha\beta\gamma\delta}D_{\alpha\beta\gamma\delta}\omega_{i\alpha}\omega_{i\beta}u_{j\gamma}u_{j\delta} \\
+&\sum_{ij}\sum_{\alpha\beta}D'_{ij\alpha\beta}\omega_{i\alpha}\omega_{i\beta}u_{j\alpha}+\sum_{i}\sum_{\alpha\beta}D''_{ij\alpha\beta}\omega_{i\alpha}u_{i\beta} \nonumber \; ,
\end{align}
where the sum runs over all the sites (perovskite cell), with $i$ and $j$ being first nearest neighbors of each other. The three terms are, respectively, the bi-quadratic (only $D_{xyxy}$ here \cite{sm}), tri-linear, and bi-linear coupling between the local mode vector ${\bf u}$ and the pseudo vector characterizing tilting ${\bm{\omega}}$. 

We then carried out Monte Carlo simulations based on the effective Hamiltonian, up to 40,000 MC sweeps at 1 K. We typically adopted a 4$\times$4$\times$2 supercell in terms of 5-atom perovskite cells, which can thus accommodate the \textit{Pbam}, \textit{R3c} and \textit{Pnma} structures. With the DFT determined coefficients, as shown by the vertical lines in Fig. \ref{fig4}, the \textit{Pbam} phase is indeed found to be the ground state, while the \textit{R3c} and \textit{Pnma} phases are 4 and 27 meV/f.u. higher in energy, respectively. This is in excellent agreement with direct DFT calculations. We also constructed, relaxed and stabilized a complex structure with a 16$\times$16$\times$2 supercell, which has a long modulation corresponding to $\bm{q}_{\text{LM}}=(2\pi/a)(3/16,3/16,0)$  \cite{note4}. This complex structure is merely 11 meV/f.u. above the \textit{Pbam} phase, indicating that  modulated structures may occur as  intermediate phase, which is reminiscent of the incommensurate structure having a similar k-point and  experimentally found in PZO under pressure \cite{burkovsky2017}. To see the effect of each type of coupling on structural stability, we individually vary the coefficients $D$ (related to $u^2\omega^2$-type of coupling), $D'$ (corresponding to the $u\omega^2$ type of couplings), and $D''$ (associated with the $u\omega$ type of couplings), and calculate the total energies (Fig. \ref{fig4}). It is important to note that each type of coupling has significant influence on one specific phase. For instance, larger magnitude of $D$  (or $D'$ ) favors the \textit{R3c} (or \textit{Pnma}) phase, as it is known before in other perovskites \cite{kornev2009,bellaiche2013}. PZO is especially interesting that the strength of the bi-linear coupling $u\omega$ is found to play an important role to render the \textit{Pbam} phase ground state, as recently proposed by Patel \textit{et al.}  \cite{patel2016}, in particular for perovskites with A-site cations that tend to move off-center, e.g., Pb$^{2+}$ or Bi$^{3+}$ with lone-pair electrons. These findings also reveal that  the $u\omega^2$ tri-linear coupling that was proposed in Refs. \cite{hlinka2014,jorge2014,prosandeev2014}  to stabilize the \textit{Pbam} structure may, in fact, not be the main contribution for PZO, as it lowers the energy of \textit{Pbam} only slightly (Fig. \ref{fig4}b) \cite{note5}.




Our calculations therefore, e.g.,  (i) emphasize the  order-disorder character (due to dynamical hopping); (ii)  point out the importance of bi-linear couplings between the local dipoles and octahedral tiltings; and (iii) suggest possible complex candidates or phenomena for the intermediate phase of PbZrO$_3$. We thus hope that our findings provide a better understanding of antiferroelectric perovskites, and suggest possible design of new antiferroelectrics having \textit{Pbam} or related structures, by searching for novel systems with large bi-linear coupling.

\begin{acknowledgments}
	We thank J. \'I\~niguez, B. Dkhil, P.-E. Janolin, R. Burkovsky and J. Hlinka for insightful discussions. B.X. thanks the support of the Air Force Office of Scientific Research under Grant No. FA9550-16-1-0065.  L.B. acknowledges the ONR Grant  No. N00014-17-1-2818. We also thank the computational support from Arkansas High Performance Computer Center at the University of Arkansas, and Penguin Computing's On-Demand HPC supported by the Cloud Pilot project.
\end{acknowledgments}



\begin{thebibliography}{61}%
	\makeatletter
	\providecommand \@ifxundefined [1]{%
		\@ifx{#1\undefined}
	}%
	\providecommand \@ifnum [1]{%
		\ifnum #1\expandafter \@firstoftwo
		\else \expandafter \@secondoftwo
		\fi
	}%
	\providecommand \@ifx [1]{%
		\ifx #1\expandafter \@firstoftwo
		\else \expandafter \@secondoftwo
		\fi
	}%
	\providecommand \natexlab [1]{#1}%
	\providecommand \enquote  [1]{``#1''}%
	\providecommand \bibnamefont  [1]{#1}%
	\providecommand \bibfnamefont [1]{#1}%
	\providecommand \citenamefont [1]{#1}%
	\providecommand \href@noop [0]{\@secondoftwo}%
	\providecommand \href [0]{\begingroup \@sanitize@url \@href}%
	\providecommand \@href[1]{\@@startlink{#1}\@@href}%
	\providecommand \@@href[1]{\endgroup#1\@@endlink}%
	\providecommand \@sanitize@url [0]{\catcode `\\12\catcode `\$12\catcode
		`\&12\catcode `\#12\catcode `\^12\catcode `\_12\catcode `\%12\relax}%
	\providecommand \@@startlink[1]{}%
	\providecommand \@@endlink[0]{}%
	\providecommand \url  [0]{\begingroup\@sanitize@url \@url }%
	\providecommand \@url [1]{\endgroup\@href {#1}{\urlprefix }}%
	\providecommand \urlprefix  [0]{URL }%
	\providecommand \Eprint [0]{\href }%
	\providecommand \doibase [0]{http://dx.doi.org/}%
	\providecommand \selectlanguage [0]{\@gobble}%
	\providecommand \bibinfo  [0]{\@secondoftwo}%
	\providecommand \bibfield  [0]{\@secondoftwo}%
	\providecommand \translation [1]{[#1]}%
	\providecommand \BibitemOpen [0]{}%
	\providecommand \bibitemStop [0]{}%
	\providecommand \bibitemNoStop [0]{.\EOS\space}%
	\providecommand \EOS [0]{\spacefactor3000\relax}%
	\providecommand \BibitemShut  [1]{\csname bibitem#1\endcsname}%
	\let\auto@bib@innerbib\@empty
	\bibitem [{\citenamefont {Kittel}(1951)}]{kittel1951}%
	\BibitemOpen
	\bibfield  {author} {\bibinfo {author} {\bibfnamefont {C.}~\bibnamefont
			{Kittel}},\ }\href@noop {} {\bibfield  {journal} {\bibinfo  {journal} {Phys.
				Rev.}\ }\textbf {\bibinfo {volume} {82}},\ \bibinfo {pages} {729} (\bibinfo
		{year} {1951})}\BibitemShut {NoStop}%
	\bibitem [{\citenamefont {Hao}(2013)}]{hao2013}%
	\BibitemOpen
	\bibfield  {author} {\bibinfo {author} {\bibfnamefont {X.}~\bibnamefont
			{Hao}},\ }\href@noop {} {\bibfield  {journal} {\bibinfo  {journal} {J. Adv.
				Dielect.}\ }\textbf {\bibinfo {volume} {3}},\ \bibinfo {pages} {1330001}
		(\bibinfo {year} {2013})}\BibitemShut {NoStop}%
	\bibitem [{\citenamefont {Rabe}(2013)}]{rabe2013}%
	\BibitemOpen
	\bibfield  {author} {\bibinfo {author} {\bibfnamefont {K.~M.}\ \bibnamefont
			{Rabe}},\ }\href@noop {} {\emph {\bibinfo {title} {Functional Metal
				Oxides}}},\ edited by\ \bibinfo {editor} {\bibfnamefont {S.~B.}\ \bibnamefont
		{Ogale}}, \bibinfo {editor} {\bibfnamefont {T.~V.}\ \bibnamefont
		{Venkatesan}}, \ and\ \bibinfo {editor} {\bibfnamefont {M.}~\bibnamefont
		{Blamire}},\ \bibinfo {number} {pp. 221-244}\ (\bibinfo  {publisher} {Wiley,
		New York},\ \bibinfo {year} {2013})\BibitemShut {NoStop}%
	\bibitem [{\citenamefont {Xu}\ \emph {et~al.}(2017)\citenamefont {Xu},
		\citenamefont {{\'I}{\~n}iguez},\ and\ \citenamefont {Bellaiche}}]{xu2017}%
	\BibitemOpen
	\bibfield  {author} {\bibinfo {author} {\bibfnamefont {B.}~\bibnamefont
			{Xu}}, \bibinfo {author} {\bibfnamefont {J.}~\bibnamefont {{\'I}{\~n}iguez}},
		\ and\ \bibinfo {author} {\bibfnamefont {L.}~\bibnamefont {Bellaiche}},\
	}\href@noop {} {\bibfield  {journal} {\bibinfo  {journal} {Nat. Commun.}\
		}\textbf {\bibinfo {volume} {8}},\ \bibinfo {pages} {15682} (\bibinfo {year}
		{2017})}\BibitemShut {NoStop}%
	\bibitem [{\citenamefont {Sawaguchi}\ \emph {et~al.}(1951)\citenamefont
		{Sawaguchi}, \citenamefont {Shirane},\ and\ \citenamefont
		{Takagi}}]{sawaguchi1951}%
	\BibitemOpen
	\bibfield  {author} {\bibinfo {author} {\bibfnamefont {E.}~\bibnamefont
			{Sawaguchi}}, \bibinfo {author} {\bibfnamefont {G.}~\bibnamefont {Shirane}},
		\ and\ \bibinfo {author} {\bibfnamefont {Y.}~\bibnamefont {Takagi}},\
	}\href@noop {} {\bibfield  {journal} {\bibinfo  {journal} {J. Phys. Soc.
				Jpn.}\ }\textbf {\bibinfo {volume} {6}},\ \bibinfo {pages} {333} (\bibinfo
		{year} {1951})}\BibitemShut {NoStop}%
	\bibitem [{\citenamefont {Shirane}\ \emph {et~al.}(1951)\citenamefont
		{Shirane}, \citenamefont {Sawaguchi},\ and\ \citenamefont
		{Takagi}}]{shirane1951}%
	\BibitemOpen
	\bibfield  {author} {\bibinfo {author} {\bibfnamefont {G.}~\bibnamefont
			{Shirane}}, \bibinfo {author} {\bibfnamefont {E.}~\bibnamefont {Sawaguchi}},
		\ and\ \bibinfo {author} {\bibfnamefont {Y.}~\bibnamefont {Takagi}},\ }\href
	{\doibase 10.1103/PhysRev.84.476} {\bibfield  {journal} {\bibinfo  {journal}
			{Phys. Rev.}\ }\textbf {\bibinfo {volume} {84}},\ \bibinfo {pages} {476}
		(\bibinfo {year} {1951})}\BibitemShut {NoStop}%
	\bibitem [{\citenamefont {Shirane}(1952)}]{shirane1952}%
	\BibitemOpen
	\bibfield  {author} {\bibinfo {author} {\bibfnamefont {G.}~\bibnamefont
			{Shirane}},\ }\href {\doibase 10.1103/PhysRev.86.219} {\bibfield  {journal}
		{\bibinfo  {journal} {Phys. Rev.}\ }\textbf {\bibinfo {volume} {86}},\
		\bibinfo {pages} {219} (\bibinfo {year} {1952})}\BibitemShut {NoStop}%
	\bibitem [{\citenamefont {Tagantsev}\ \emph {et~al.}(2013)\citenamefont
		{Tagantsev}, \citenamefont {Vaideeswaran}, \citenamefont {Vakhrushev},
		\citenamefont {Filimonov}, \citenamefont {Burkovsky}, \citenamefont
		{Shaganov}, \citenamefont {Andronikova}, \citenamefont {Rudskoy},
		\citenamefont {Baron}, \citenamefont {Uchiyama} \emph
		{et~al.}}]{tagantsev2013}%
	\BibitemOpen
	\bibfield  {author} {\bibinfo {author} {\bibfnamefont {A.}~\bibnamefont
			{Tagantsev}}, \bibinfo {author} {\bibfnamefont {K.}~\bibnamefont
			{Vaideeswaran}}, \bibinfo {author} {\bibfnamefont {S.}~\bibnamefont
			{Vakhrushev}}, \bibinfo {author} {\bibfnamefont {A.}~\bibnamefont
			{Filimonov}}, \bibinfo {author} {\bibfnamefont {R.}~\bibnamefont
			{Burkovsky}}, \bibinfo {author} {\bibfnamefont {A.}~\bibnamefont {Shaganov}},
		\bibinfo {author} {\bibfnamefont {D.}~\bibnamefont {Andronikova}}, \bibinfo
		{author} {\bibfnamefont {A.}~\bibnamefont {Rudskoy}}, \bibinfo {author}
		{\bibfnamefont {A.}~\bibnamefont {Baron}}, \bibinfo {author} {\bibfnamefont
			{H.}~\bibnamefont {Uchiyama}},  \emph {et~al.},\ }\href@noop {} {\bibfield
		{journal} {\bibinfo  {journal} {Nat. Commun.}\ }\textbf {\bibinfo {volume}
			{4}},\ \bibinfo {pages} {ncomms3229} (\bibinfo {year} {2013})}\BibitemShut
	{NoStop}%
	\bibitem [{\citenamefont {\'I\~niguez}\ \emph {et~al.}(2014)\citenamefont
		{\'I\~niguez}, \citenamefont {Stengel}, \citenamefont {Prosandeev},\ and\
		\citenamefont {Bellaiche}}]{jorge2014}%
	\BibitemOpen
	\bibfield  {author} {\bibinfo {author} {\bibfnamefont {J.}~\bibnamefont
			{\'I\~niguez}}, \bibinfo {author} {\bibfnamefont {M.}~\bibnamefont
			{Stengel}}, \bibinfo {author} {\bibfnamefont {S.}~\bibnamefont {Prosandeev}},
		\ and\ \bibinfo {author} {\bibfnamefont {L.}~\bibnamefont {Bellaiche}},\
	}\href@noop {} {\bibfield  {journal} {\bibinfo  {journal} {Phys. Rev. B}\
		}\textbf {\bibinfo {volume} {90}},\ \bibinfo {pages} {220103} (\bibinfo
		{year} {2014})}\BibitemShut {NoStop}%
	\bibitem [{\citenamefont {Hlinka}\ \emph {et~al.}(2014)\citenamefont {Hlinka},
		\citenamefont {Ostapchuk}, \citenamefont {Buixaderas}, \citenamefont
		{Kadlec}, \citenamefont {Kuzel}, \citenamefont {Gregora}, \citenamefont
		{Kroupa}, \citenamefont {Savinov}, \citenamefont {Klic}, \citenamefont
		{Drahokoupil}, \citenamefont {Etxebarria},\ and\ \citenamefont
		{Dec}}]{hlinka2014}%
	\BibitemOpen
	\bibfield  {author} {\bibinfo {author} {\bibfnamefont {J.}~\bibnamefont
			{Hlinka}}, \bibinfo {author} {\bibfnamefont {T.}~\bibnamefont {Ostapchuk}},
		\bibinfo {author} {\bibfnamefont {E.}~\bibnamefont {Buixaderas}}, \bibinfo
		{author} {\bibfnamefont {C.}~\bibnamefont {Kadlec}}, \bibinfo {author}
		{\bibfnamefont {P.}~\bibnamefont {Kuzel}}, \bibinfo {author} {\bibfnamefont
			{I.}~\bibnamefont {Gregora}}, \bibinfo {author} {\bibfnamefont
			{J.}~\bibnamefont {Kroupa}}, \bibinfo {author} {\bibfnamefont
			{M.}~\bibnamefont {Savinov}}, \bibinfo {author} {\bibfnamefont
			{A.}~\bibnamefont {Klic}}, \bibinfo {author} {\bibfnamefont {J.}~\bibnamefont
			{Drahokoupil}}, \bibinfo {author} {\bibfnamefont {I.}~\bibnamefont
			{Etxebarria}}, \ and\ \bibinfo {author} {\bibfnamefont {J.}~\bibnamefont
			{Dec}},\ }\href@noop {} {\bibfield  {journal} {\bibinfo  {journal} {Phys.
				Rev. Lett.}\ }\textbf {\bibinfo {volume} {112}},\ \bibinfo {pages} {197601}
		(\bibinfo {year} {2014})}\BibitemShut {NoStop}%
	\bibitem [{\citenamefont {Fthenakis}\ and\ \citenamefont
		{Ponomareva}(2017)}]{fthenakis2017}%
	\BibitemOpen
	\bibfield  {author} {\bibinfo {author} {\bibfnamefont {Z.~G.}\ \bibnamefont
			{Fthenakis}}\ and\ \bibinfo {author} {\bibfnamefont {I.}~\bibnamefont
			{Ponomareva}},\ }\href@noop {} {\bibfield  {journal} {\bibinfo  {journal}
			{Phys. Rev. B}\ }\textbf {\bibinfo {volume} {96}},\ \bibinfo {pages} {184110}
		(\bibinfo {year} {2017})}\BibitemShut {NoStop}%
	\bibitem [{\citenamefont {Vales-Castro}\ \emph {et~al.}(2018)\citenamefont
		{Vales-Castro}, \citenamefont {Roleder}, \citenamefont {Zhao}, \citenamefont
		{Li}, \citenamefont {Kajewski},\ and\ \citenamefont {Catalan}}]{vales2018}%
	\BibitemOpen
	\bibfield  {author} {\bibinfo {author} {\bibfnamefont {P.}~\bibnamefont
			{Vales-Castro}}, \bibinfo {author} {\bibfnamefont {K.}~\bibnamefont
			{Roleder}}, \bibinfo {author} {\bibfnamefont {L.}~\bibnamefont {Zhao}},
		\bibinfo {author} {\bibfnamefont {J.-F.}\ \bibnamefont {Li}}, \bibinfo
		{author} {\bibfnamefont {D.}~\bibnamefont {Kajewski}}, \ and\ \bibinfo
		{author} {\bibfnamefont {G.}~\bibnamefont {Catalan}},\ }\href@noop {}
	{\bibfield  {journal} {\bibinfo  {journal} {Appl. Phys. Lett.}\ }\textbf
		{\bibinfo {volume} {113}},\ \bibinfo {pages} {132903} (\bibinfo {year}
		{2018})}\BibitemShut {NoStop}%
	\bibitem [{\citenamefont {Ostapchuk}\ \emph {et~al.}(2001)\citenamefont
		{Ostapchuk}, \citenamefont {Petzelt}, \citenamefont {Zelezny}, \citenamefont
		{Kamba}, \citenamefont {Bovtun}, \citenamefont {Porokhonskyy}, \citenamefont
		{Pashkin}, \citenamefont {Kuzel}, \citenamefont {Glinchuk}, \citenamefont
		{Bykov} \emph {et~al.}}]{ostapchuk2001}%
	\BibitemOpen
	\bibfield  {author} {\bibinfo {author} {\bibfnamefont {T.}~\bibnamefont
			{Ostapchuk}}, \bibinfo {author} {\bibfnamefont {J.}~\bibnamefont {Petzelt}},
		\bibinfo {author} {\bibfnamefont {V.}~\bibnamefont {Zelezny}}, \bibinfo
		{author} {\bibfnamefont {S.}~\bibnamefont {Kamba}}, \bibinfo {author}
		{\bibfnamefont {V.}~\bibnamefont {Bovtun}}, \bibinfo {author} {\bibfnamefont
			{V.}~\bibnamefont {Porokhonskyy}}, \bibinfo {author} {\bibfnamefont
			{A.}~\bibnamefont {Pashkin}}, \bibinfo {author} {\bibfnamefont
			{P.}~\bibnamefont {Kuzel}}, \bibinfo {author} {\bibfnamefont
			{M.}~\bibnamefont {Glinchuk}}, \bibinfo {author} {\bibfnamefont
			{I.}~\bibnamefont {Bykov}},  \emph {et~al.},\ }\href@noop {} {\bibfield
		{journal} {\bibinfo  {journal} {J. Phys. Condens. Matter}\ }\textbf {\bibinfo
			{volume} {13}},\ \bibinfo {pages} {2677} (\bibinfo {year}
		{2001})}\BibitemShut {NoStop}%
	\bibitem [{\citenamefont {Ko}\ \emph {et~al.}(2013)\citenamefont {Ko},
		\citenamefont {G\'orny}, \citenamefont {Majchrowski}, \citenamefont
		{Roleder},\ and\ \citenamefont {Bussmann-Holder}}]{ko2013}%
	\BibitemOpen
	\bibfield  {author} {\bibinfo {author} {\bibfnamefont {J.-H.}\ \bibnamefont
			{Ko}}, \bibinfo {author} {\bibfnamefont {M.}~\bibnamefont {G\'orny}},
		\bibinfo {author} {\bibfnamefont {A.}~\bibnamefont {Majchrowski}}, \bibinfo
		{author} {\bibfnamefont {K.}~\bibnamefont {Roleder}}, \ and\ \bibinfo
		{author} {\bibfnamefont {A.}~\bibnamefont {Bussmann-Holder}},\ }\href
	{\doibase 10.1103/PhysRevB.87.184110} {\bibfield  {journal} {\bibinfo
			{journal} {Phys. Rev. B}\ }\textbf {\bibinfo {volume} {87}},\ \bibinfo
		{pages} {184110} (\bibinfo {year} {2013})}\BibitemShut {NoStop}%
	\bibitem [{\citenamefont {Pa\'sciak}\ \emph {et~al.}(2015)\citenamefont
		{Pa\'sciak}, \citenamefont {Welberry}, \citenamefont {Heerdegen},
		\citenamefont {Laguta}, \citenamefont {Ostapchuk}, \citenamefont {Leoni},\
		and\ \citenamefont {Hlinka}}]{pasciak2015}%
	\BibitemOpen
	\bibfield  {author} {\bibinfo {author} {\bibfnamefont {M.}~\bibnamefont
			{Pa\'sciak}}, \bibinfo {author} {\bibfnamefont {T.}~\bibnamefont {Welberry}},
		\bibinfo {author} {\bibfnamefont {A.}~\bibnamefont {Heerdegen}}, \bibinfo
		{author} {\bibfnamefont {V.}~\bibnamefont {Laguta}}, \bibinfo {author}
		{\bibfnamefont {T.}~\bibnamefont {Ostapchuk}}, \bibinfo {author}
		{\bibfnamefont {S.}~\bibnamefont {Leoni}}, \ and\ \bibinfo {author}
		{\bibfnamefont {J.}~\bibnamefont {Hlinka}},\ }\href {\doibase
		10.1080/01411594.2014.981266} {\bibfield  {journal} {\bibinfo  {journal}
			{Phase Transitions}\ }\textbf {\bibinfo {volume} {88}},\ \bibinfo {pages}
		{273} (\bibinfo {year} {2015})}\BibitemShut {NoStop}%
	\bibitem [{\citenamefont {Zhang}\ \emph {et~al.}(2015)\citenamefont {Zhang},
		\citenamefont {Pa{\'{s}}ciak}, \citenamefont {Glazer}, \citenamefont
		{Hlinka}, \citenamefont {Gutmann}, \citenamefont {Sparkes}, \citenamefont
		{Welberry}, \citenamefont {Majchrowski}, \citenamefont {Roleder},
		\citenamefont {Xie},\ and\ \citenamefont {Ye}}]{zhang2015}%
	\BibitemOpen
	\bibfield  {author} {\bibinfo {author} {\bibfnamefont {N.}~\bibnamefont
			{Zhang}}, \bibinfo {author} {\bibfnamefont {M.}~\bibnamefont
			{Pa{\'{s}}ciak}}, \bibinfo {author} {\bibfnamefont {A.~M.}\ \bibnamefont
			{Glazer}}, \bibinfo {author} {\bibfnamefont {J.}~\bibnamefont {Hlinka}},
		\bibinfo {author} {\bibfnamefont {M.}~\bibnamefont {Gutmann}}, \bibinfo
		{author} {\bibfnamefont {H.~A.}\ \bibnamefont {Sparkes}}, \bibinfo {author}
		{\bibfnamefont {T.~R.}\ \bibnamefont {Welberry}}, \bibinfo {author}
		{\bibfnamefont {A.}~\bibnamefont {Majchrowski}}, \bibinfo {author}
		{\bibfnamefont {K.}~\bibnamefont {Roleder}}, \bibinfo {author} {\bibfnamefont
			{Y.}~\bibnamefont {Xie}}, \ and\ \bibinfo {author} {\bibfnamefont {Z.-G.}\
			\bibnamefont {Ye}},\ }\href {\doibase 10.1107/S1600576715017069} {\bibfield
		{journal} {\bibinfo  {journal} {J. Appl. Crystallogr.}\ }\textbf {\bibinfo
			{volume} {48}},\ \bibinfo {pages} {1637} (\bibinfo {year}
		{2015})}\BibitemShut {NoStop}%
	\bibitem [{\citenamefont {Klarbring}\ and\ \citenamefont
		{Simak}(2018)}]{klarbring2018}%
	\BibitemOpen
	\bibfield  {author} {\bibinfo {author} {\bibfnamefont {J.}~\bibnamefont
			{Klarbring}}\ and\ \bibinfo {author} {\bibfnamefont {S.~I.}\ \bibnamefont
			{Simak}},\ }\href {\doibase 10.1103/PhysRevB.97.024108} {\bibfield  {journal}
		{\bibinfo  {journal} {Phys. Rev. B}\ }\textbf {\bibinfo {volume} {97}},\
		\bibinfo {pages} {024108} (\bibinfo {year} {2018})}\BibitemShut {NoStop}%
	\bibitem [{\citenamefont {Klarbring}(2019)}]{klarbring2019}%
	\BibitemOpen
	\bibfield  {author} {\bibinfo {author} {\bibfnamefont {J.}~\bibnamefont
			{Klarbring}},\ }\href {\doibase 10.1103/PhysRevB.99.104105} {\bibfield
		{journal} {\bibinfo  {journal} {Phys. Rev. B}\ }\textbf {\bibinfo {volume}
			{99}},\ \bibinfo {pages} {104105} (\bibinfo {year} {2019})}\BibitemShut
	{NoStop}%
	\bibitem [{\citenamefont {Reyes-Lillo}\ and\ \citenamefont
		{Rabe}(2013)}]{sebastian2013}%
	\BibitemOpen
	\bibfield  {author} {\bibinfo {author} {\bibfnamefont {S.~E.}\ \bibnamefont
			{Reyes-Lillo}}\ and\ \bibinfo {author} {\bibfnamefont {K.~M.}\ \bibnamefont
			{Rabe}},\ }\href {\doibase 10.1103/PhysRevB.88.180102} {\bibfield  {journal}
		{\bibinfo  {journal} {Phys. Rev. B}\ }\textbf {\bibinfo {volume} {88}},\
		\bibinfo {pages} {180102} (\bibinfo {year} {2013})}\BibitemShut {NoStop}%
	\bibitem [{\citenamefont {Patel}\ \emph {et~al.}(2016)\citenamefont {Patel},
		\citenamefont {Prosandeev}, \citenamefont {Yang}, \citenamefont {Xu},
		\citenamefont {\'I\~niguez},\ and\ \citenamefont {Bellaiche}}]{patel2016}%
	\BibitemOpen
	\bibfield  {author} {\bibinfo {author} {\bibfnamefont {K.}~\bibnamefont
			{Patel}}, \bibinfo {author} {\bibfnamefont {S.}~\bibnamefont {Prosandeev}},
		\bibinfo {author} {\bibfnamefont {Y.}~\bibnamefont {Yang}}, \bibinfo {author}
		{\bibfnamefont {B.}~\bibnamefont {Xu}}, \bibinfo {author} {\bibfnamefont
			{J.}~\bibnamefont {\'I\~niguez}}, \ and\ \bibinfo {author} {\bibfnamefont
			{L.}~\bibnamefont {Bellaiche}},\ }\href {\doibase 10.1103/PhysRevB.94.054107}
	{\bibfield  {journal} {\bibinfo  {journal} {Phys. Rev. B}\ }\textbf {\bibinfo
			{volume} {94}},\ \bibinfo {pages} {054107} (\bibinfo {year}
		{2016})}\BibitemShut {NoStop}%
	\bibitem [{\citenamefont {Tennery}(1965)}]{tennery1965}%
	\BibitemOpen
	\bibfield  {author} {\bibinfo {author} {\bibfnamefont {V.~J.}\ \bibnamefont
			{Tennery}},\ }\href@noop {} {\bibfield  {journal} {\bibinfo  {journal} {J.
				Electrochem. Soc.}\ }\textbf {\bibinfo {volume} {112}},\ \bibinfo {pages}
		{1117} (\bibinfo {year} {1965})}\BibitemShut {NoStop}%
	\bibitem [{\citenamefont {Tennery}(1966)}]{tennery1966}%
	\BibitemOpen
	\bibfield  {author} {\bibinfo {author} {\bibfnamefont {V.~J.}\ \bibnamefont
			{Tennery}},\ }\href@noop {} {\bibfield  {journal} {\bibinfo  {journal} {J.
				Am. Ceram. Soc.}\ }\textbf {\bibinfo {volume} {49}},\ \bibinfo {pages} {483}
		(\bibinfo {year} {1966})}\BibitemShut {NoStop}%
	\bibitem [{\citenamefont {Goulpeau}(1967)}]{goulpeau1967}%
	\BibitemOpen
	\bibfield  {author} {\bibinfo {author} {\bibfnamefont {L.}~\bibnamefont
			{Goulpeau}},\ }\href@noop {} {\bibfield  {journal} {\bibinfo  {journal} {Sov.
				Phys.-Solid State}\ }\textbf {\bibinfo {volume} {8}},\ \bibinfo {pages}
		{1970} (\bibinfo {year} {1967})}\BibitemShut {NoStop}%
	\bibitem [{\citenamefont {Fujishita}(1992)}]{fujishita1992}%
	\BibitemOpen
	\bibfield  {author} {\bibinfo {author} {\bibfnamefont {H.}~\bibnamefont
			{Fujishita}},\ }\href@noop {} {\bibfield  {journal} {\bibinfo  {journal} {J.
				Phys. Soc. Jpn.}\ }\textbf {\bibinfo {volume} {61}},\ \bibinfo {pages} {3606}
		(\bibinfo {year} {1992})}\BibitemShut {NoStop}%
	\bibitem [{\citenamefont {Liu}(2018)}]{liu2018}%
	\BibitemOpen
	\bibfield  {author} {\bibinfo {author} {\bibfnamefont {H.}~\bibnamefont
			{Liu}},\ }\href@noop {} {\bibfield  {journal} {\bibinfo  {journal} {J. Am.
				Ceram. Soc.}\ }\textbf {\bibinfo {volume} {101}},\ \bibinfo {pages} {5281}
		(\bibinfo {year} {2018})}\BibitemShut {NoStop}%
	\bibitem [{\citenamefont {Kresse}\ and\ \citenamefont
		{Joubert}(1999)}]{kresse1999}%
	\BibitemOpen
	\bibfield  {author} {\bibinfo {author} {\bibfnamefont {G.}~\bibnamefont
			{Kresse}}\ and\ \bibinfo {author} {\bibfnamefont {D.}~\bibnamefont
			{Joubert}},\ }\href {\doibase 10.1103/PhysRevB.59.1758} {\bibfield  {journal}
		{\bibinfo  {journal} {Phys. Rev. B}\ }\textbf {\bibinfo {volume} {59}},\
		\bibinfo {pages} {1758} (\bibinfo {year} {1999})}\BibitemShut {NoStop}%
	\bibitem [{\citenamefont {Hellman}\ \emph {et~al.}(2011)\citenamefont
		{Hellman}, \citenamefont {Abrikosov},\ and\ \citenamefont
		{Simak}}]{hellman2011}%
	\BibitemOpen
	\bibfield  {author} {\bibinfo {author} {\bibfnamefont {O.}~\bibnamefont
			{Hellman}}, \bibinfo {author} {\bibfnamefont {I.~A.}\ \bibnamefont
			{Abrikosov}}, \ and\ \bibinfo {author} {\bibfnamefont {S.~I.}\ \bibnamefont
			{Simak}},\ }\href@noop {} {\bibfield  {journal} {\bibinfo  {journal} {Phys.
				Rev. B}\ }\textbf {\bibinfo {volume} {84}},\ \bibinfo {pages} {180301}
		(\bibinfo {year} {2011})}\BibitemShut {NoStop}%
	\bibitem [{\citenamefont {Hellman}\ and\ \citenamefont
		{Abrikosov}(2013)}]{hellman2013}%
	\BibitemOpen
	\bibfield  {author} {\bibinfo {author} {\bibfnamefont {O.}~\bibnamefont
			{Hellman}}\ and\ \bibinfo {author} {\bibfnamefont {I.~A.}\ \bibnamefont
			{Abrikosov}},\ }\href@noop {} {\bibfield  {journal} {\bibinfo  {journal}
			{Phys. Rev. B}\ }\textbf {\bibinfo {volume} {88}},\ \bibinfo {pages} {144301}
		(\bibinfo {year} {2013})}\BibitemShut {NoStop}%
	\bibitem [{\citenamefont {Hellman}\ \emph {et~al.}(2013)\citenamefont
		{Hellman}, \citenamefont {Steneteg}, \citenamefont {Abrikosov},\ and\
		\citenamefont {Simak}}]{hellman2013_2}%
	\BibitemOpen
	\bibfield  {author} {\bibinfo {author} {\bibfnamefont {O.}~\bibnamefont
			{Hellman}}, \bibinfo {author} {\bibfnamefont {P.}~\bibnamefont {Steneteg}},
		\bibinfo {author} {\bibfnamefont {I.~A.}\ \bibnamefont {Abrikosov}}, \ and\
		\bibinfo {author} {\bibfnamefont {S.~I.}\ \bibnamefont {Simak}},\ }\href@noop
	{} {\bibfield  {journal} {\bibinfo  {journal} {Phys. Rev. B}\ }\textbf
		{\bibinfo {volume} {87}},\ \bibinfo {pages} {104111} (\bibinfo {year}
		{2013})}\BibitemShut {NoStop}%
	\bibitem [{\citenamefont {Henkelman}\ \emph {et~al.}(2000)\citenamefont
		{Henkelman}, \citenamefont {Uberuaga},\ and\ \citenamefont
		{J{\'o}nsson}}]{henkelman2000}%
	\BibitemOpen
	\bibfield  {author} {\bibinfo {author} {\bibfnamefont {G.}~\bibnamefont
			{Henkelman}}, \bibinfo {author} {\bibfnamefont {B.~P.}\ \bibnamefont
			{Uberuaga}}, \ and\ \bibinfo {author} {\bibfnamefont {H.}~\bibnamefont
			{J{\'o}nsson}},\ }\href@noop {} {\bibfield  {journal} {\bibinfo  {journal}
			{J. Chem. Phys.}\ }\textbf {\bibinfo {volume} {113}},\ \bibinfo {pages}
		{9901} (\bibinfo {year} {2000})}\BibitemShut {NoStop}%
	\bibitem [{sm()}]{sm}%
	\BibitemOpen
	\href@noop {} {}\bibinfo {note} {See Supplemental Material for (1) details of
		the computational methods; (2) comparison of the effect of MD and stochastic
		sampling on the phonon dispersion; (3) Additional analysis from MD
		simulations, which includes Refs.
		\cite{kresse1999,perdew2008,faye_phd,hellman2011,hellman2013,hellman2013_2,phonopy,zhong1994,zhong1995,kornev2006,mani2015prb,mani2015,fthenakis2017}}\BibitemShut
	{NoStop}%
	\bibitem [{\citenamefont {Ghosez}\ \emph {et~al.}(1999)\citenamefont {Ghosez},
		\citenamefont {Cockayne}, \citenamefont {Waghmare},\ and\ \citenamefont
		{Rabe}}]{ghosez1999}%
	\BibitemOpen
	\bibfield  {author} {\bibinfo {author} {\bibfnamefont {P.}~\bibnamefont
			{Ghosez}}, \bibinfo {author} {\bibfnamefont {E.}~\bibnamefont {Cockayne}},
		\bibinfo {author} {\bibfnamefont {U.~V.}\ \bibnamefont {Waghmare}}, \ and\
		\bibinfo {author} {\bibfnamefont {K.~M.}\ \bibnamefont {Rabe}},\ }\href@noop
	{} {\bibfield  {journal} {\bibinfo  {journal} {Phys. Rev. B}\ }\textbf
		{\bibinfo {volume} {60}},\ \bibinfo {pages} {836} (\bibinfo {year}
		{1999})}\BibitemShut {NoStop}%
	\bibitem [{\citenamefont {Cockayne}\ and\ \citenamefont
		{Rabe}(2000)}]{cockayne2000}%
	\BibitemOpen
	\bibfield  {author} {\bibinfo {author} {\bibfnamefont {E.}~\bibnamefont
			{Cockayne}}\ and\ \bibinfo {author} {\bibfnamefont {K.}~\bibnamefont
			{Rabe}},\ }\href@noop {} {\bibfield  {journal} {\bibinfo  {journal} {J. Phys.
				Chem. Solids}\ }\textbf {\bibinfo {volume} {61}},\ \bibinfo {pages} {305}
		(\bibinfo {year} {2000})}\BibitemShut {NoStop}%
	\bibitem [{\citenamefont {Leung}\ \emph {et~al.}(2002)\citenamefont {Leung},
		\citenamefont {Cockayne},\ and\ \citenamefont {Wright}}]{leung2002}%
	\BibitemOpen
	\bibfield  {author} {\bibinfo {author} {\bibfnamefont {K.}~\bibnamefont
			{Leung}}, \bibinfo {author} {\bibfnamefont {E.}~\bibnamefont {Cockayne}}, \
		and\ \bibinfo {author} {\bibfnamefont {A.~F.}\ \bibnamefont {Wright}},\
	}\href {\doibase 10.1103/PhysRevB.65.214111} {\bibfield  {journal} {\bibinfo
			{journal} {Phys. Rev. B}\ }\textbf {\bibinfo {volume} {65}},\ \bibinfo
		{pages} {214111} (\bibinfo {year} {2002})}\BibitemShut {NoStop}%
	\bibitem [{\citenamefont {Leung}(2003)}]{leung2003}%
	\BibitemOpen
	\bibfield  {author} {\bibinfo {author} {\bibfnamefont {K.}~\bibnamefont
			{Leung}},\ }\href {\doibase 10.1103/PhysRevB.67.104108} {\bibfield  {journal}
		{\bibinfo  {journal} {Phys. Rev. B}\ }\textbf {\bibinfo {volume} {67}},\
		\bibinfo {pages} {104108} (\bibinfo {year} {2003})}\BibitemShut {NoStop}%
	\bibitem [{not({\natexlab{a}})}]{note1}%
	\BibitemOpen
	\href@noop {} {} \bibinfo {note} {Below 450 K, if the
		initial atomic positions are \textit{Pbam}, the structure remains
		\textit{Pbam}; whereas if we start with cubic positions, the average atomic
		positions deviate to a \textit{R3c} symmetry.}\BibitemShut {Stop}%
	\bibitem [{\citenamefont {Samara}(1970)}]{samara1970}%
	\BibitemOpen
	\bibfield  {author} {\bibinfo {author} {\bibfnamefont {G.~A.}\ \bibnamefont
			{Samara}},\ }\href {\doibase 10.1103/PhysRevB.1.3777} {\bibfield  {journal}
		{\bibinfo  {journal} {Phys. Rev. B}\ }\textbf {\bibinfo {volume} {1}},\
		\bibinfo {pages} {3777} (\bibinfo {year} {1970})}\BibitemShut {NoStop}%
	\bibitem [{\citenamefont {Whatmore}\ and\ \citenamefont
		{Glazer}(1979)}]{whatmore1979}%
	\BibitemOpen
	\bibfield  {author} {\bibinfo {author} {\bibfnamefont {R.}~\bibnamefont
			{Whatmore}}\ and\ \bibinfo {author} {\bibfnamefont {A.}~\bibnamefont
			{Glazer}},\ }\href@noop {} {\bibfield  {journal} {\bibinfo  {journal} {J.
				Phys. C: Solid State Phys.}\ }\textbf {\bibinfo {volume} {12}},\ \bibinfo
		{pages} {1505} (\bibinfo {year} {1979})}\BibitemShut {NoStop}%
	\bibitem [{\citenamefont {Fujishita}\ \emph {et~al.}(2003)\citenamefont
		{Fujishita}, \citenamefont {Ishikawa}, \citenamefont {Tanaka}, \citenamefont
		{Ogawaguchi},\ and\ \citenamefont {Katano}}]{fujishita2003}%
	\BibitemOpen
	\bibfield  {author} {\bibinfo {author} {\bibfnamefont {H.}~\bibnamefont
			{Fujishita}}, \bibinfo {author} {\bibfnamefont {Y.}~\bibnamefont {Ishikawa}},
		\bibinfo {author} {\bibfnamefont {S.}~\bibnamefont {Tanaka}}, \bibinfo
		{author} {\bibfnamefont {A.}~\bibnamefont {Ogawaguchi}}, \ and\ \bibinfo
		{author} {\bibfnamefont {S.}~\bibnamefont {Katano}},\ }\href@noop {}
	{\bibfield  {journal} {\bibinfo  {journal} {J. Phys. Soc. Jpn.}\ }\textbf
		{\bibinfo {volume} {72}},\ \bibinfo {pages} {1426} (\bibinfo {year}
		{2003})}\BibitemShut {NoStop}%
	\bibitem [{not({\natexlab{b}})}]{note3}%
	\BibitemOpen
	\href@noop {} {} \bibinfo {note} {phonon dispersion
		calculated based on the stochastic sampling at high temperature has soft
		phonons, which resembles the DFPT phonon at 0 K (see Supplemental
		Materials).}\BibitemShut {Stop}%
	\bibitem [{\citenamefont {Orobengoa}\ \emph {et~al.}(2009)\citenamefont
		{Orobengoa}, \citenamefont {Capillas}, \citenamefont {Aroyo},\ and\
		\citenamefont {Perez-Mato}}]{orobengoa2009}%
	\BibitemOpen
	\bibfield  {author} {\bibinfo {author} {\bibfnamefont {D.}~\bibnamefont
			{Orobengoa}}, \bibinfo {author} {\bibfnamefont {C.}~\bibnamefont {Capillas}},
		\bibinfo {author} {\bibfnamefont {M.~I.}\ \bibnamefont {Aroyo}}, \ and\
		\bibinfo {author} {\bibfnamefont {J.~M.}\ \bibnamefont {Perez-Mato}},\
	}\href@noop {} {\bibfield  {journal} {\bibinfo  {journal} {J. Appl.
				Crystallogr.}\ }\textbf {\bibinfo {volume} {42}},\ \bibinfo {pages} {820}
		(\bibinfo {year} {2009})}\BibitemShut {NoStop}%
	\bibitem [{\citenamefont {Perez-Mato}\ \emph {et~al.}(2010)\citenamefont
		{Perez-Mato}, \citenamefont {Orobengoa},\ and\ \citenamefont
		{Aroyo}}]{perez2010}%
	\BibitemOpen
	\bibfield  {author} {\bibinfo {author} {\bibfnamefont {J.}~\bibnamefont
			{Perez-Mato}}, \bibinfo {author} {\bibfnamefont {D.}~\bibnamefont
			{Orobengoa}}, \ and\ \bibinfo {author} {\bibfnamefont {M.}~\bibnamefont
			{Aroyo}},\ }\href@noop {} {\bibfield  {journal} {\bibinfo  {journal} {Acta
				Cryst. A}\ }\textbf {\bibinfo {volume} {66}},\ \bibinfo {pages} {558}
		(\bibinfo {year} {2010})}\BibitemShut {NoStop}%
	\bibitem [{not({\natexlab{c}})}]{note2}%
	\BibitemOpen
	\href@noop {} {} \bibinfo {note} {For all the PES
		calculations, the same cubic cell is adopted, while the atomic distortions of
		the $Pbam$, $R3c$, and $Pnma$ structures are incorporated. The effect caused
		by the difference in lattice parameters should be secondary. In addition,
		minor distortions in \textit{Pbam}, i.e., $R^-_4$, $X^+_1$, and $M^-_5$
		modes, are not considered here because of their much smaller amplitude. These
		factors give rise to an energy being 12.7 meV/f.u. higher in the minima of
		PES than that of the fully relaxed \textit{Pbam} phase.}\BibitemShut {Stop}%
	\bibitem [{\citenamefont {Ayyub}\ \emph {et~al.}(1998)\citenamefont {Ayyub},
		\citenamefont {Chattopadhyay}, \citenamefont {Pinto},\ and\ \citenamefont
		{Multani}}]{ayyub1998}%
	\BibitemOpen
	\bibfield  {author} {\bibinfo {author} {\bibfnamefont {P.}~\bibnamefont
			{Ayyub}}, \bibinfo {author} {\bibfnamefont {S.}~\bibnamefont
			{Chattopadhyay}}, \bibinfo {author} {\bibfnamefont {R.}~\bibnamefont
			{Pinto}}, \ and\ \bibinfo {author} {\bibfnamefont {M.~S.}\ \bibnamefont
			{Multani}},\ }\href@noop {} {\bibfield  {journal} {\bibinfo  {journal} {Phys.
				Rev. B}\ }\textbf {\bibinfo {volume} {57}},\ \bibinfo {pages} {R5559}
		(\bibinfo {year} {1998})}\BibitemShut {NoStop}%
	\bibitem [{\citenamefont {Boldyreva}\ \emph {et~al.}(2007)\citenamefont
		{Boldyreva}, \citenamefont {Pintilie}, \citenamefont {Lotnyk}, \citenamefont
		{Misirlioglu}, \citenamefont {Alexe},\ and\ \citenamefont
		{Hesse}}]{boldyreva2007}%
	\BibitemOpen
	\bibfield  {author} {\bibinfo {author} {\bibfnamefont {K.}~\bibnamefont
			{Boldyreva}}, \bibinfo {author} {\bibfnamefont {L.}~\bibnamefont {Pintilie}},
		\bibinfo {author} {\bibfnamefont {A.}~\bibnamefont {Lotnyk}}, \bibinfo
		{author} {\bibfnamefont {I.~B.}\ \bibnamefont {Misirlioglu}}, \bibinfo
		{author} {\bibfnamefont {M.}~\bibnamefont {Alexe}}, \ and\ \bibinfo {author}
		{\bibfnamefont {D.}~\bibnamefont {Hesse}},\ }\href@noop {} {\bibfield
		{journal} {\bibinfo  {journal} {Appl. Phys. Lett.}\ }\textbf {\bibinfo
			{volume} {91}},\ \bibinfo {pages} {122915} (\bibinfo {year}
		{2007})}\BibitemShut {NoStop}%
	\bibitem [{\citenamefont {Mani}\ \emph
		{et~al.}(2015{\natexlab{a}})\citenamefont {Mani}, \citenamefont {Chang},
		\citenamefont {Lisenkov},\ and\ \citenamefont {Ponomareva}}]{mani2015}%
	\BibitemOpen
	\bibfield  {author} {\bibinfo {author} {\bibfnamefont {B.~K.}\ \bibnamefont
			{Mani}}, \bibinfo {author} {\bibfnamefont {C.-M.}\ \bibnamefont {Chang}},
		\bibinfo {author} {\bibfnamefont {S.}~\bibnamefont {Lisenkov}}, \ and\
		\bibinfo {author} {\bibfnamefont {I.}~\bibnamefont {Ponomareva}},\ }\href
	{\doibase 10.1103/PhysRevLett.115.097601} {\bibfield  {journal} {\bibinfo
			{journal} {Phys. Rev. Lett.}\ }\textbf {\bibinfo {volume} {115}},\ \bibinfo
		{pages} {097601} (\bibinfo {year} {2015}{\natexlab{a}})}\BibitemShut
	{NoStop}%
	\bibitem [{\citenamefont {Woodward}\ \emph {et~al.}(2005)\citenamefont
		{Woodward}, \citenamefont {Knudsen},\ and\ \citenamefont
		{Reaney}}]{woodward2005}%
	\BibitemOpen
	\bibfield  {author} {\bibinfo {author} {\bibfnamefont {D.~I.}\ \bibnamefont
			{Woodward}}, \bibinfo {author} {\bibfnamefont {J.}~\bibnamefont {Knudsen}}, \
		and\ \bibinfo {author} {\bibfnamefont {I.~M.}\ \bibnamefont {Reaney}},\
	}\href {\doibase 10.1103/PhysRevB.72.104110} {\bibfield  {journal} {\bibinfo
			{journal} {Phys. Rev. B}\ }\textbf {\bibinfo {volume} {72}},\ \bibinfo
		{pages} {104110} (\bibinfo {year} {2005})}\BibitemShut {NoStop}%
	\bibitem [{\citenamefont {Scott}\ and\ \citenamefont
		{Burns}(1972)}]{scott1972}%
	\BibitemOpen
	\bibfield  {author} {\bibinfo {author} {\bibfnamefont {B.}~\bibnamefont
			{Scott}}\ and\ \bibinfo {author} {\bibfnamefont {G.}~\bibnamefont {Burns}},\
	}\href@noop {} {\bibfield  {journal} {\bibinfo  {journal} {J. Am. Ceram.
				Soc.}\ }\textbf {\bibinfo {volume} {55}},\ \bibinfo {pages} {331} (\bibinfo
		{year} {1972})}\BibitemShut {NoStop}%
	\bibitem [{not({\natexlab{d}})}]{note4}%
	\BibitemOpen
	\href@noop {} {} \bibinfo {note} {To use this effective
		Hamiltonian in order to simulate finite-temperature properties, a potential
		issue needs to be solved first, that is unphysical AFD patterns corresponding
		to the $k$-point $\Gamma$ or $X$ due to the bi-linear coupling should be
		avoided.}\BibitemShut {Stop}%
	\bibitem [{\citenamefont {Burkovsky}\ \emph {et~al.}(2017)\citenamefont
		{Burkovsky}, \citenamefont {Bronwald}, \citenamefont {Andronikova},
		\citenamefont {Wehinger}, \citenamefont {Krisch}, \citenamefont {Jacobs},
		\citenamefont {Gambetti}, \citenamefont {Roleder}, \citenamefont
		{Majchrowski}, \citenamefont {Filimonov} \emph {et~al.}}]{burkovsky2017}%
	\BibitemOpen
	\bibfield  {author} {\bibinfo {author} {\bibfnamefont {R.}~\bibnamefont
			{Burkovsky}}, \bibinfo {author} {\bibfnamefont {I.}~\bibnamefont {Bronwald}},
		\bibinfo {author} {\bibfnamefont {D.}~\bibnamefont {Andronikova}}, \bibinfo
		{author} {\bibfnamefont {B.}~\bibnamefont {Wehinger}}, \bibinfo {author}
		{\bibfnamefont {M.}~\bibnamefont {Krisch}}, \bibinfo {author} {\bibfnamefont
			{J.}~\bibnamefont {Jacobs}}, \bibinfo {author} {\bibfnamefont
			{D.}~\bibnamefont {Gambetti}}, \bibinfo {author} {\bibfnamefont
			{K.}~\bibnamefont {Roleder}}, \bibinfo {author} {\bibfnamefont
			{A.}~\bibnamefont {Majchrowski}}, \bibinfo {author} {\bibfnamefont
			{A.}~\bibnamefont {Filimonov}},  \emph {et~al.},\ }\href@noop {} {\bibfield
		{journal} {\bibinfo  {journal} {Sci. Rep.}\ }\textbf {\bibinfo {volume}
			{7}},\ \bibinfo {pages} {41512} (\bibinfo {year} {2017})}\BibitemShut
	{NoStop}%
	\bibitem [{\citenamefont {Kornev}\ and\ \citenamefont
		{Bellaiche}(2009)}]{kornev2009}%
	\BibitemOpen
	\bibfield  {author} {\bibinfo {author} {\bibfnamefont {I.~A.}\ \bibnamefont
			{Kornev}}\ and\ \bibinfo {author} {\bibfnamefont {L.}~\bibnamefont
			{Bellaiche}},\ }\href {\doibase 10.1103/PhysRevB.79.100105} {\bibfield
		{journal} {\bibinfo  {journal} {Phys. Rev. B}\ }\textbf {\bibinfo {volume}
			{79}},\ \bibinfo {pages} {100105} (\bibinfo {year} {2009})}\BibitemShut
	{NoStop}%
	\bibitem [{\citenamefont {Bellaiche}\ and\ \citenamefont
		{\'I\~niguez}(2013)}]{bellaiche2013}%
	\BibitemOpen
	\bibfield  {author} {\bibinfo {author} {\bibfnamefont {L.}~\bibnamefont
			{Bellaiche}}\ and\ \bibinfo {author} {\bibfnamefont {J.}~\bibnamefont
			{\'I\~niguez}},\ }\href@noop {} {\bibfield  {journal} {\bibinfo  {journal}
			{Phys. Rev. B}\ }\textbf {\bibinfo {volume} {88}},\ \bibinfo {pages} {014104}
		(\bibinfo {year} {2013})}\BibitemShut {NoStop}%
	\bibitem [{\citenamefont {Prosandeev}\ \emph {et~al.}(2014)\citenamefont
		{Prosandeev}, \citenamefont {Xu}, \citenamefont {Faye}, \citenamefont {Duan},
		\citenamefont {Liu}, \citenamefont {Dkhil}, \citenamefont {Janolin},
		\citenamefont {\'I\~niguez},\ and\ \citenamefont
		{Bellaiche}}]{prosandeev2014}%
	\BibitemOpen
	\bibfield  {author} {\bibinfo {author} {\bibfnamefont {S.}~\bibnamefont
			{Prosandeev}}, \bibinfo {author} {\bibfnamefont {C.}~\bibnamefont {Xu}},
		\bibinfo {author} {\bibfnamefont {R.}~\bibnamefont {Faye}}, \bibinfo {author}
		{\bibfnamefont {W.}~\bibnamefont {Duan}}, \bibinfo {author} {\bibfnamefont
			{H.}~\bibnamefont {Liu}}, \bibinfo {author} {\bibfnamefont {B.}~\bibnamefont
			{Dkhil}}, \bibinfo {author} {\bibfnamefont {P.-E.}\ \bibnamefont {Janolin}},
		\bibinfo {author} {\bibfnamefont {J.}~\bibnamefont {\'I\~niguez}}, \ and\
		\bibinfo {author} {\bibfnamefont {L.}~\bibnamefont {Bellaiche}},\ }\href
	{\doibase 10.1103/PhysRevB.89.214111} {\bibfield  {journal} {\bibinfo
			{journal} {Phys. Rev. B}\ }\textbf {\bibinfo {volume} {89}},\ \bibinfo
		{pages} {214111} (\bibinfo {year} {2014})}\BibitemShut {NoStop}%
	\bibitem [{not({\natexlab{e}})}]{note5}%
	\BibitemOpen
	\href@noop {} {} \bibinfo {note} {Note that such result is
		consistent with the aforementioned small amplitude of the S$_2$ mode found by
		DFT calculations in the $Pbam$ ground state of PZO, emphasizing that the
		tri-linear coupling involving the $R^-_5$, $\Sigma_2$, and S$_2$ modes is
		relatively small.}\BibitemShut {Stop}%
	\bibitem [{\citenamefont {Perdew}\ \emph {et~al.}(2008)\citenamefont {Perdew},
		\citenamefont {Ruzsinszky}, \citenamefont {Csonka}, \citenamefont {Vydrov},
		\citenamefont {Scuseria}, \citenamefont {Constantin}, \citenamefont {Zhou},\
		and\ \citenamefont {Burke}}]{perdew2008}%
	\BibitemOpen
	\bibfield  {author} {\bibinfo {author} {\bibfnamefont {J.~P.}\ \bibnamefont
			{Perdew}}, \bibinfo {author} {\bibfnamefont {A.}~\bibnamefont {Ruzsinszky}},
		\bibinfo {author} {\bibfnamefont {G.~I.}\ \bibnamefont {Csonka}}, \bibinfo
		{author} {\bibfnamefont {O.~A.}\ \bibnamefont {Vydrov}}, \bibinfo {author}
		{\bibfnamefont {G.~E.}\ \bibnamefont {Scuseria}}, \bibinfo {author}
		{\bibfnamefont {L.~A.}\ \bibnamefont {Constantin}}, \bibinfo {author}
		{\bibfnamefont {X.}~\bibnamefont {Zhou}}, \ and\ \bibinfo {author}
		{\bibfnamefont {K.}~\bibnamefont {Burke}},\ }\href {\doibase
		10.1103/PhysRevLett.100.136406} {\bibfield  {journal} {\bibinfo  {journal}
			{Phys. Rev. Lett.}\ }\textbf {\bibinfo {volume} {100}},\ \bibinfo {pages}
		{136406} (\bibinfo {year} {2008})}\BibitemShut {NoStop}%
	\bibitem [{\citenamefont {Faye}(2014)}]{faye_phd}%
	\BibitemOpen
	\bibfield  {author} {\bibinfo {author} {\bibfnamefont {R.}~\bibnamefont
			{Faye}},\ }\emph {\bibinfo {title} {Structures et propriétés d'un
			antiferroélectrique modèle}},\ \href@noop {} {Ph.D. thesis},\ \bibinfo
	{school} {Ecole centrale de Paris} (\bibinfo {year} {2014})\BibitemShut
	{NoStop}%
	\bibitem [{\citenamefont {Togo}\ and\ \citenamefont {Tanaka}(2015)}]{phonopy}%
	\BibitemOpen
	\bibfield  {author} {\bibinfo {author} {\bibfnamefont {A.}~\bibnamefont
			{Togo}}\ and\ \bibinfo {author} {\bibfnamefont {I.}~\bibnamefont {Tanaka}},\
	}\href@noop {} {\bibfield  {journal} {\bibinfo  {journal} {Scr. Mater.}\
		}\textbf {\bibinfo {volume} {108}},\ \bibinfo {pages} {1} (\bibinfo {year}
		{2015})}\BibitemShut {NoStop}%
	\bibitem [{\citenamefont {Zhong}\ \emph {et~al.}(1994)\citenamefont {Zhong},
		\citenamefont {Vanderbilt},\ and\ \citenamefont {Rabe}}]{zhong1994}%
	\BibitemOpen
	\bibfield  {author} {\bibinfo {author} {\bibfnamefont {W.}~\bibnamefont
			{Zhong}}, \bibinfo {author} {\bibfnamefont {D.}~\bibnamefont {Vanderbilt}}, \
		and\ \bibinfo {author} {\bibfnamefont {K.~M.}\ \bibnamefont {Rabe}},\
	}\href@noop {} {\bibfield  {journal} {\bibinfo  {journal} {Phys. Rev. Lett.}\
		}\textbf {\bibinfo {volume} {73}},\ \bibinfo {pages} {1861} (\bibinfo {year}
		{1994})}\BibitemShut {NoStop}%
	\bibitem [{\citenamefont {Zhong}\ \emph {et~al.}(1995)\citenamefont {Zhong},
		\citenamefont {Vanderbilt},\ and\ \citenamefont {Rabe}}]{zhong1995}%
	\BibitemOpen
	\bibfield  {author} {\bibinfo {author} {\bibfnamefont {W.}~\bibnamefont
			{Zhong}}, \bibinfo {author} {\bibfnamefont {D.}~\bibnamefont {Vanderbilt}}, \
		and\ \bibinfo {author} {\bibfnamefont {K.~M.}\ \bibnamefont {Rabe}},\
	}\href@noop {} {\bibfield  {journal} {\bibinfo  {journal} {Phys. Rev. B}\
		}\textbf {\bibinfo {volume} {52}},\ \bibinfo {pages} {6301} (\bibinfo {year}
		{1995})}\BibitemShut {NoStop}%
	\bibitem [{\citenamefont {Kornev}\ \emph {et~al.}(2006)\citenamefont {Kornev},
		\citenamefont {Bellaiche}, \citenamefont {Janolin}, \citenamefont {Dkhil},\
		and\ \citenamefont {Suard}}]{kornev2006}%
	\BibitemOpen
	\bibfield  {author} {\bibinfo {author} {\bibfnamefont {I.~A.}\ \bibnamefont
			{Kornev}}, \bibinfo {author} {\bibfnamefont {L.}~\bibnamefont {Bellaiche}},
		\bibinfo {author} {\bibfnamefont {P.-E.}\ \bibnamefont {Janolin}}, \bibinfo
		{author} {\bibfnamefont {B.}~\bibnamefont {Dkhil}}, \ and\ \bibinfo {author}
		{\bibfnamefont {E.}~\bibnamefont {Suard}},\ }\href@noop {} {\bibfield
		{journal} {\bibinfo  {journal} {Phys. Rev. Lett.}\ }\textbf {\bibinfo
			{volume} {97}},\ \bibinfo {pages} {157601} (\bibinfo {year}
		{2006})}\BibitemShut {NoStop}%
	\bibitem [{\citenamefont {Mani}\ \emph
		{et~al.}(2015{\natexlab{b}})\citenamefont {Mani}, \citenamefont {Lisenkov},\
		and\ \citenamefont {Ponomareva}}]{mani2015prb}%
	\BibitemOpen
	\bibfield  {author} {\bibinfo {author} {\bibfnamefont {B.~K.}\ \bibnamefont
			{Mani}}, \bibinfo {author} {\bibfnamefont {S.}~\bibnamefont {Lisenkov}}, \
		and\ \bibinfo {author} {\bibfnamefont {I.}~\bibnamefont {Ponomareva}},\
	}\href {\doibase 10.1103/PhysRevB.91.134112} {\bibfield  {journal} {\bibinfo
			{journal} {Phys. Rev. B}\ }\textbf {\bibinfo {volume} {91}},\ \bibinfo
		{pages} {134112} (\bibinfo {year} {2015}{\natexlab{b}})}\BibitemShut
	{NoStop}%
\end{thebibliography}

%

\end{document}